\documentclass[12pt]{article}
\usepackage{amsmath}
\usepackage{amssymb}
\usepackage{color}
\usepackage{epsfig}

\makeatletter \@addtoreset{equation}{section} \makeatother
\makeatletter \@addtoreset{figure}{section} \makeatother

\def\IC{\mathbb{C}}

\def\IZ{\mathbb{Z}}

\def\CL{{\cal L}}

\newcommand{\be}{\begin{equation}}
\newcommand{\ee}{\end{equation}}
\newcommand{\bea}{\begin{eqnarray}}
\newcommand{\eea}{\end{eqnarray}}
\newcommand{\ba}{\begin{array}}

\def\ba{{\bar a}}
\def\be{{\bar e}}

\begin{document}
\begin{titlepage}
\vfill
\begin{flushright}
{\tt\normalsize KIAS-P12030}\\

\end{flushright}
\vfill
\begin{center}
{\Large\bf
Quiver Invariants from Intrinsic Higgs States}

\vskip 1cm
Seung-Joo Lee,\footnote{\tt s.lee@kias.re.kr}
Zhao-Long Wang,\footnote{\tt zlwang@kias.re.kr}
and Piljin Yi\footnote{\tt piljin@kias.re.kr}

\vskip 5mm
{\it School of Physics, Korea Institute
for Advanced Study, Seoul 130-722, Korea}

\end{center}
\vfill

\begin{abstract}
\noindent
In study of four-dimensional BPS states, quiver quantum
mechanics plays a central role. The Coulomb phases capture
the multi-centered nature of such states, and are
well understood in the context of wall-crossing. The Higgs
phases are given typically by F-term-induced complete intersections  in the
ambient D-term-induced toric varieties, and the ground states can be
far more numerous than the Coulomb
phase counterparts. We observe that the Higgs phase BPS
states are naturally and geometrically grouped into two parts, with
one part given by the pulled-back cohomology from the
D-term-induced ambient space. We propose that
these pulled-back states are in one-to-one correspondence
with the Coulomb phase states. This also
leads us to conjecture
that the index associated with the rest, intrinsic to the Higgs phase,
is a fundamental invariant of quivers, independent of branches.
For simple circular quivers, these intrinsic Higgs states
belong to the middle cohomology and thus are all angular momentum
singlets, supporting the single-center black hole interpretation.

\end{abstract}

\vfill
\end{titlepage}

\tableofcontents\newpage
\renewcommand{\thefootnote}{\#\arabic{footnote}}
\setcounter{footnote}{0}

\parskip 0.2 cm

\section{Conjectures}

There are many stringy realizations of $D=4$, $N=2$ supersymmetric theories,
and accordingly, many realizations of BPS states \cite{Prasad:1975kr}
thereof. Among the latter, the most powerful and all-encompassing
picture appears to be the quiver realization due to Denef~\cite{Denef:2002ru}.
The simplest way to motivate the quiver quantum mechanics is to consider the
BPS states as wrapped D3 branes in Calabi-Yau-compactified
type IIB theories. The BPS condition then translates to Special
Lagrange (SL) property of the 3-cycle wrapped by D3; when
the 3-cycle is a sum of simpler SL 3-cycles,
this induces a low energy quantum mechanics of the constituent
BPS particles of quiver type with four supercharges.

The Coulomb phase of such quiver quantum mechanics has been
studied in many guises, but in the end is always related to
wall-crossing phenomena \cite{Seiberg:1994rs,Ferrari:1996sv}. Natural description of a BPS state
in this phase is as a multi-center bound state where centers of
relatively non-local charges are held fixed relative to each
other via balance of classical forces~\cite{Lee:1998nv}.
Celebrated wall-crossing phenomena are simple consequences of this
multi-center picture, when the size of a bound state diverges as one
approaches a wall in the vacuum moduli space or the parameter
space~\cite{Bak:1999da,Gauntlett:1999vc,Gauntlett:2000ks,Denef:2000nb}.\footnote{See Ref.~\cite{Weinberg:2006rq}
for a general review from field theory perspective.}
Ground state counting in the Coulomb phase has been understood
extensively~\cite{Stern:2000ie,Denef:2002ru,Denef:2007vg,deBoer:2008zn,Manschot:2010qz,Pioline:2011gf,Manschot:2011xc,Lee:2011ph,Kim:2011sc},
from what the right index problem is to how such
indices are related to those of $D=4$, $N=2$ field theory.
The resulting wall-crossing formula~\cite{Manschot:2010qz,Manschot:2011xc,Kim:2011sc}
has been proved~\cite{Sen:2011aa} to be consistent with
abstract wall-crossing formulae~\cite{KS,Gaiotto:2008cd,Joyce:2008pc} as well,
including that of Kontsevich and Soibelman.

The Higgs phase is of entirely different character and is
represented by a complete intersection, call it $M$, in an
ambient projective variety, call it $X$. This is so because
D-term conditions associated with the gauge symmetries of
the quiver produce a projective toric variety, X, whereas
the F-term conditions reduce the phase to the zero-locus of
a set of polynomials, say $\partial W=0$, when a superpotential $W$
exists. We mean, by Higgs phase, this complete intersection manifold
$M$ assuming a generic superpotential $W$.
Then the supersymmetric ground states are represented naturally by
the cohomology ring, $H(M)$, which gives a simple way to count the index thereof,
\begin{equation}
\chi(M)=\sum_n (-1)^n\;{\rm dim}H^n(M)\,,
\end{equation}
as the alternative sum of cohomology group dimensions.

For simple quivers, such as loop-less ones, the Higgs phase counting is found to agree with
the Coulomb phase. For more generic cases, however, in particular
for those quivers that contain a loop and also allow, in the Coulomb phase,
certain ``scaling" solutions associated with the loops,
it is known that the Higgs counting is often much larger than the Coulomb one, and can even be exponentially so
at that. As the Coulomb phase is supposed to capture multi-center states,
it is then natural to expect that these extra states in Higgs phase,
which we will call ``intrinsic Higgs states", are intrinsic to a single-centered
black hole and the entropy thereof.

No matter what the correct physical interpretations of these numerous states are,
however, we must first understand exactly how the intrinsic Higgs states are
characterized in the Higgs phase of quiver quantum
mechanics.\footnote{ One might be tempted to search for
another type of index that only captures the intrinsic Higgs states. An obvious
candidate, the signature of $M$, can be seen not to work at all when
complex dimension of $M$ is odd, since it vanishes then, and typically
too small in the even cases also.} For this, we note that the Higgs
phase $M$ as a complete intersection inside a projective toric space $X$
suggests an interesting dichotomy of $H(M)$ as follows,
\begin{equation}
H(M)=i^*_M(H(X))\oplus \left[H(M)/i^*_M(H(X))\right] ,
\end{equation}
where $i^*_M$ is the pull-back onto $M$ using the embedding map $i_M: M\hookrightarrow X$.
We may further define $\chi( i^*_M(H(X)))$ as the usual alternating sum
of Betti numbers. Note that $\chi( i^*_M(H(X)))={\rm dim}\;i^*_M(H(X))$ if
the ring $H(X)$ only consists of even cohomology groups,
as is the case for the type of quivers we study in this note. Hence, we use the two alternatively.

With this  dichotomy of Higgs phase ground states, we
now conjecture the following:
\begin{itemize}

\item
The pull-back $i^*_M(H(X))$ of the ambient cohomology is in
one-to-one correspondence with the Coulomb phase states.
In particular,
$$\Omega_{\rm Coulomb}={\rm dim}\;i^*_M(H(X))\,,$$
where $\Omega_{\rm Coulomb}$ is the Coulomb phase index, and hence,
according to our terminology, the states {\it not} in $i^*_M(H(X))$
are intrinsic Higgs states.

\item
These intrinsic Higgs states in $H(M)/i^*_M(H(X))$ are inherent
to the quiver quantum mechanics in that their counting given by
$$\chi(M)- \chi(i^*_M(H(X)))\,,$$
does not change as we cross walls by adjusting Fayet-Illiopoulos (FI) constants.

\end{itemize}
We will be checking the conjectures
for the simplest class of quivers made of a circular loop.
Simple generalizations, such as connecting one or more
such circular quivers with trees, follow immediately
as long as we do not create new loops in the process.
The second conjecture appears to be quite a nontrivial statement,
especially when taken as a purely geometric statement.
It would be most interesting to understand it from  mathematical
perspectives, but this is beyond the scope of this note.

We stated the conjectures as two separate ones, but
they are of course linked to each other, once we believe that
the underlying wall-crossing behavior is entirely captured
by the Coulomb phase ground states. The latter has been
convincingly demonstrated \cite{Sen:2011aa} when the quiver has no
loops,\footnote{Most rigorous check of which was for
the Coulomb phase wall-crossing against Kontsevich-Soibelman
algebra when all charge vectors
involved are on a single plane passing through origin.} but
general expectation is that the same will hold for quivers
with loops, since the fundamental physics underlying wall-crossing
is in the multi-center nature, as captured faithfully by the
Coulomb phase.

There is one immediate evidence for the first conjecture from
the angular momentum consideration. As we will see in section 3,
for simple circular quivers, the intrinsic Higgs
phase states are all angular momentum singlets. This fact alone goes a long way to support
the conjecture since, if the single-center black hole interpretation for
intrinsic Higgs states is
correct, we naturally expect no angular momentum associated with
them.\footnote{Vanishing angular momentum for single-center states has been
conjectured and checked  for $N=4$ black holes recently
\cite{Sen:2009vz,Sen:2010mz}.}

In the next section, we review geometry of the Higgs branches for a quiver
with a single circular loop, and comment on the Coulomb counterpart.
In particular, we identify the ambient space $X$ as a product of projective
spaces, $\mathbb{CP}^n$'s, and present basic topology of the Higgs phase
$M$ as a complete intersection manifold embedded therein.
In section 3, we explore both $H(M)$ and $i^*_M(H(X))$.
In section 4, we take the case of simplest circular quiver, with three nodes, and show validity of
these conjectures explicitly.
In section 5,  we make  further checks of the conjectures
numerically for quivers with four or more nodes. In section 6, we  close with comments.

\section{Higgs Phases of  Quivers with a Loop}

\begin{figure}[!h]
\centering
\includegraphics[width=11cm]{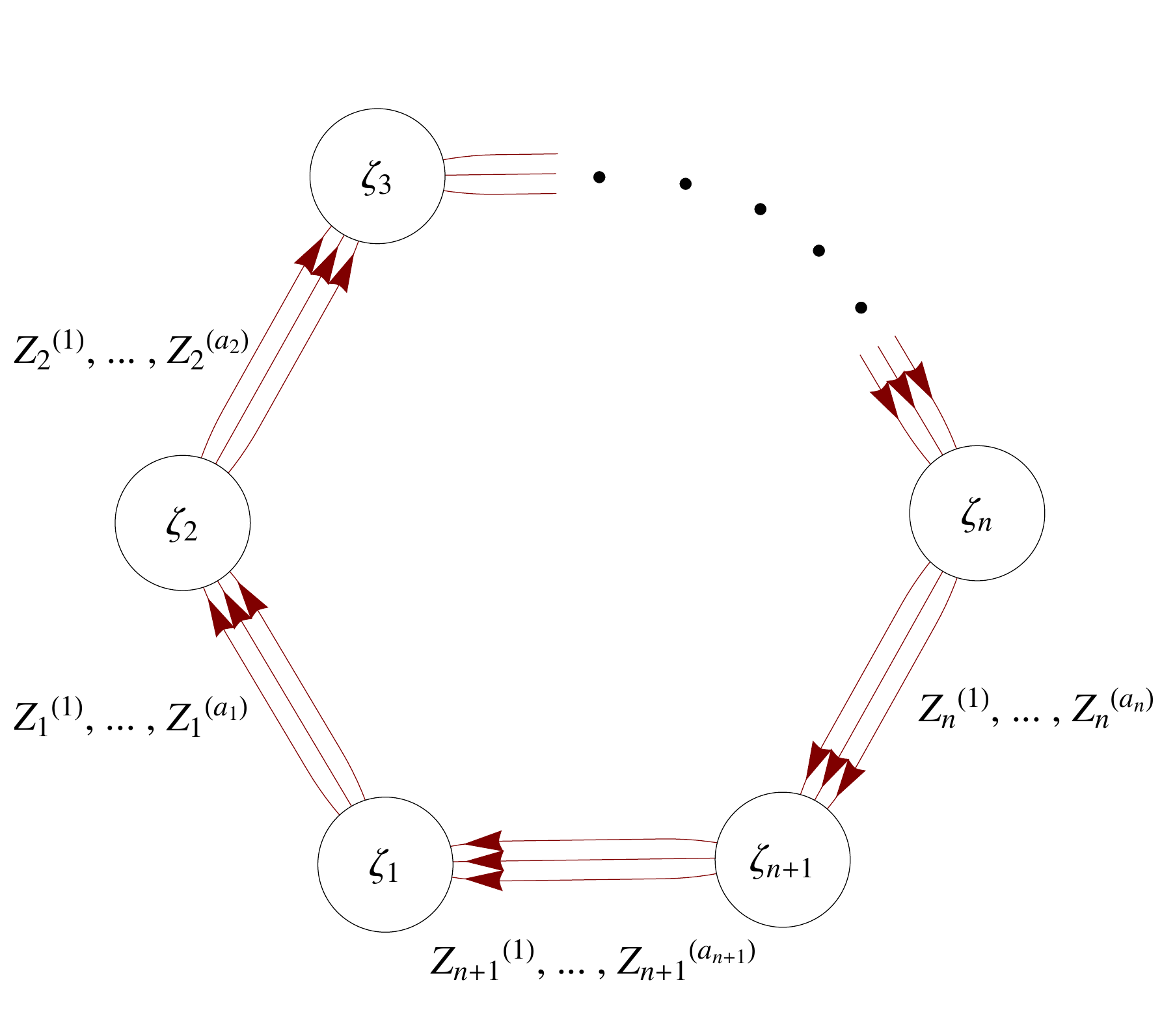}
\parbox{6in}{\caption{\small A cyclic $(n+1)$-gon quiver consists
of $n+1$ nodes, cyclically connected by directed arrows. Associated
to each node is a $U(1)$ gauge group, whose FI constant is denoted
by $\zeta_i$, and the $a_i$ arrows from the $i$-th node to the $(i+1)$-th
node correspond to the $a_i$ bifundamental fields $Z_i = (Z_i^{(1)},
\cdots, Z_i^{(a_i)})$.}\label{fig:quiver}}
\end{figure}

Let us start with a cyclic $(n+1)$-gon quiver, and denote the bifundamental
fields ${Z}_{i,i+1}$ by ${Z}_{i}$ for $i=1, \dots, n$, and ${Z}_{n+1,1}$ by ${Z}_{n+1}$.
For each node $i$, there are $a_i$ arrows from the $i$-th node to the $(i+1)$-th node,
where all the linking numbers $a_i$ are positive. This means that $Z_i$ are actually $a_i$-dimensional complex vectors,
\begin{equation}
Z_i = (Z_i^{(1)}, \cdots, Z_i^{(a_i)}) \ ,
\end{equation}
as depicted by Figure~\ref{fig:quiver}.
Because the quiver has a loop, with the linking numbers of the same sign,
we should expect a generic superpotential of the type,
\begin{equation}
W({Z}_1,{Z}_2,\cdots,{Z}_{n+1})=\sum_{\beta_1=1}^{a_1}\cdots
\sum_{\beta_{n+1}=1}^{a_{n+1}} c_{\beta_1\beta_2\cdots \beta_{n+1}}
{Z}^{(\beta_1)}_1{Z}^{(\beta_2)}_2\cdots{Z}^{(\beta_{n+1})}_{n+1}\,,
\end{equation}
whose F-term supersymmetric vacuum conditions are
\begin{equation}
\partial_{{Z}^{(\beta_i)}_i}W=0~~~\,(\beta_i=1,2,3,\cdots,a_i)\,,
\end{equation}
for $i=1,\cdots,n+1$.
We show that the solutions to $\partial W=0$ split into branches,
where one of the ${Z}_{i}$ complex vectors is identically zero.

Let us first show
that there is no solution to F-term conditions with all the complex vectors $Z_i$ nontrivial.
If there were such a solution,
it has to be a discrete solution since the number of F-term equations
equals the total number of complex variables $Z$'s. (We always assume that
the coefficients in $W$ are generic, so the algebraic equations $\partial W=0$
are also generic.) However, the above F-term conditions have $n+1$
scaling symmetries under $Z_i\rightarrow \lambda_iZ_i$
for any complex numbers $\lambda_i$, and one can actually generate $n+1$
complex dimensional family of solutions. This contradicts the
expected discreteness of the solution, so we cannot generically
expect to find solutions of this type.

The preceding argument may look rather prohibiting, but it can be evaded
easily by setting $Z_j=0$ for some $j$. Because of the homogeneous form of
$\partial W$'s, $Z_j=0$
renders all $\partial_{Z_{i\neq j}}W$ to vanish identically, leaving behind
only $a_j$ number of equations, $\partial_{Z_j}W=0$. In this situation,
the number of remaining variables $Z_{i\neq j}$ are typically much larger
than the surviving F-term conditions. Therefore, we conclude that, for generic
$W$, F-term conditions have to be solved by setting one of $Z_i$ to be
identically zero.

This key observation also helps us to solve D-term conditions
almost trivially. With $U(1)^{n+1}$ gauge groups,
of which a trace part $U(1)$ decouples, we have $n$-independent
D-term conditions
\begin{eqnarray}
|Z_{n+1}|^2-|Z_1|^2&=&\zeta_1\,,\\ \nonumber
|Z_1|^2-|Z_2|^2&=&\zeta_2\,,\\ \nonumber
|Z_2|^2-|Z_3|^2&=&\zeta_3\,,\\ \nonumber
&\vdots& \\ \nonumber
|Z_n|^2-|Z_{n+1}|^2&=&\zeta_{n+1}\,,\\ \nonumber
\end{eqnarray}
which require as usual
\begin{eqnarray}
\zeta_1+\cdots+\zeta_{n+1}=0\,.
\end{eqnarray}
Thanks to the above F-term discussion, we learned that
the Higgs phase has $n+1$ generic branches with one of $Z_i=0$
identically. The $k$-th branch is realized when
\begin{equation}
\sum_{i= I}^{k} \zeta_{i}> 0\ ,\quad \sum_{i= k+1}^{ J}\zeta_i<0
\end{equation}
for consecutive and mutually exclusive sets of $I$'s and $J$'s,
where the cyclic nature of the indices are understood. In this
$k$-th branch, ${Z}_{k}=0$,
and the remaining D-term conditions are solved entirely by
$$
X_k={\mathbb{CP}}^{a_1-1}\times\cdots\times {\mathbb{CP}}^{a_{k-1}-1}
\times {\mathbb{CP}}^{a_{k+1}-1}\times\cdots\times {\mathbb{CP}}^{a_{n+1}-1}\,.
$$
At the boundary of this branch, one or more of $\mathbb{CP}$'s get
squashed to zero size, and wall-crossing may occur.

Combining the two lines of thoughts, we conclude that the $k$-th  Higgs
branch, $M_k$, that emerges in a particular domain of FI constants,
has the form of a complete intersection,
\begin{equation}
M_k=X_k\biggl\vert_{\partial_{{Z}_k}W=0} \ ,
\end{equation}
given by the zero-locus of the $a_k$ F-term conditions;
All other F-term conditions are trivially met by  $Z_k=0$.

Since these phases are selected by a choice of domain in the FI constant space, each
Higgs branch has a Coulomb counterpart,  determined as a
subspace of $\mathbb{R}^{3(n+1)}$ by 1-loop zero energy conditions,
\begin{eqnarray}\label{Coulomb}
\frac{a_{n+1}}{|\vec x_{n+1}-\vec x_1|}-\frac{a_{1}}{|\vec x_{1}-\vec x_2|} &=&\zeta_1\,,\\ \nonumber
\frac{a_{1}}{|\vec x_{1}-\vec x_2|}-\frac{a_2}{|\vec x_{2}-\vec x_3|} &=&\zeta_2\,,\\ \nonumber
&\vdots& \\ \nonumber
\frac{a_{n}}{|\vec x_{n}-\vec x_{n+1}|}-\frac{a_{n+1}}{|\vec x_{n+1}-\vec x_1|} &=&\zeta_{n+1} \ ,
\end{eqnarray}
with the same  sign choices on FI constants. $\vec x_i$
represents the position of the $i$-th charge center in real
space $\mathbb{R}^3$. In this note, the corresponding Coulomb
phase index will be denoted by $\Omega_{\rm Coulomb}$, with appropriate
labels for the choice of branch.

{}From the Coulomb phase perspective, this quiver falls into two
big regimes, depending on the linking numbers $a_i$. In one
regime, so-called non-scaling regime, one finds that classical
Coulomb vacuum manifolds are such that $ |\vec x_i-\vec x_{i+1}|$'s
are all fixed. In the other, so-called
scaling regime, however, one finds classical configurations of arbitrarily
short distances among the charge centers. The criteria between the
two take a geometric character. The scaling regime occurs
when the $a_i$'s are such that they can be lengths of a
geometric $(n+1)$-gon in $\mathbb{R}^3$, i.e.,
\begin{equation}
a_j < \sum_{i\neq j} a_i
\end{equation}
for all $j$. The Coulomb index is expected to agree
with Higgs phase Euler number in the non-scaling cases,
but known to be typically smaller in the scaling cases.

\section{Basics on $H(M)$ and $i^*_M(H(X))$ for Cyclic  $(n+1)$-Gons} \label{sec3}

Let us characterize the pull-back of the ambient cohomology  for a
cyclic $(n+1)$-gon, and, as before, denote the linking numbers
as $(a_1,a_2,\cdots,a_{n+1})$, all positive. We may take, without loss of
generality, the Higgs
branch where $(n+1)$-th bifundamentals are forced to vanish. Thus, the ambient toric space is simply
\begin{equation}
X_{n+1}={\mathbb{CP}}^{a_1-1}\times {\mathbb{CP}}^{a_2-1}\times\cdots\times {\mathbb{CP}}^{a_{n}-1}\,.
\end{equation}
The superpotential $W\sim Z_1Z_2\cdots Z_{n+1}$ then yields total of
$a_{n+1}$ number of F-term conditions. This defines the Higgs
phase $M_{n+1}$, where the subscript emphasizes  the fact that
this manifold differs as we choose different branches of the
Higgs phase. The complex dimension of $M_{n+1}$ is then
\begin{equation}
d_{n+1}=a_1+a_2+\cdots+a_n-a_{n+1}-n\,.
\end{equation}
The Higgs phase index is counted by the Euler number
which is computed as integral of the top Chern class of $M_{n+1}$.
Now, the adjunction formula relates the Chern class of $M_{n+1}$
to that of the ambient space $X_{n+1}$, leading to
\begin{equation}
c(M_{n+1})=\frac{(1+J_1)^{a_1}(1+J_2)^{a_2}\cdots (1+J_n)^{a_n}}{(1+J_1+J_2+\cdots +J_n)^{a_{n+1}}}\,,
\end{equation}
where $J_i$'s are the K\"ahler class of the ${\mathbb{CP}}^{a_i-1}$ factor.
Evaluation of $\chi(M_{n+1})$ proceeds equally simply in the
ambient space $X_{n+1}$ as
\begin{equation}\label{Euler}
\chi(M_{n+1})=
\int_{X_{n+1}}\frac{(1+J_1)^{a_1}(1+J_2)^{a_2}\cdots
 (1+J_n)^{a_n}(J_1+J_2+\cdots +J_n)^{a_{n+1}}}{(1+J_1+J_2+\cdots +J_n)^{a_{n+1}}}\,,
\end{equation}
which is the same as extracting the coefficient of the top form
$J_1^{a_1-1}\cdots J_n^{a_n-1}$ of the integrand. Alternatively
and more explicitly, the Euler number can also be shown to be
\begin{eqnarray}\label{Euler2}
\chi(M_{n+1})&=&a_1a_2\cdots a_n-\int_0^\infty ds\; e^{-s}L_{a_1-1}^1(s)L_{a_2-1}^1(s)\cdots L_{a_{n+1}-1}^1(s)
\cr&=&a_1a_2\cdots a_n \\ \nonumber\
&&- \sum_{s_1=0}^{a_1-1} \sum_{s_2=0}^{a_2-1}\cdots\sum_{s_{n+1}=0}^{a_{n+1}-1}
(s_1+s_2+\cdots+s_{n+1})!\prod_{i=1}^{n+1}{a_i \choose s_i+1}\frac{(-1)^{s_i}}{s_i!}\,.
\end{eqnarray}
with Laguerre polynomials $L_{a_i-1}^1$,
following the same procedure as in \cite{Denef:2007vg}.
All of these apply to $M_k$ straightforwardly.

As we are splitting the cohomology into two parts, one of which
comes from pull-back of $H(X)$, the Lefschetz  hyperplane theorem
(see appendix A) comes in handy. When used iteratively, the theorem
implies for the circular quivers that the pull-back of $H(X)$ is
isomorphic to $H(M)$ for the lower-half cohomologies up
to $H^{d-1}$, where $d$ is the complex dimension
of $M$. Also, the map is injective for $H^d$.
Combined with the Poincar\'{e} duality, this tells us that
\begin{equation}
H^n(M)\simeq H^{2d-n}(M)\simeq H^n(X)  \ , \qquad n < d \ ,
\end{equation}
or equivalently,
\begin{equation}
H(M)=i^*_M(H(X))\oplus [H^{d}(M)/i^*_{M}(H^{d}(X))]\,.
\end{equation}
Thus, we also learned that the intrinsic Higgs states all belong to the middle
cohomology of $M$.

In turn, this  translates physically to the statement that
all intrinsic Higgs states are angular momentum
singlets, as  $(n-d)/2$ can be understood as the helicity; since $M$
is a K\"ahler manifold, the following natural actions on $n$-forms
\begin{equation}
L_3=(n-d)/2\ , \quad L_+=J\wedge  \ , \quad L_-=J \lrcorner~\,\,\,,
\end{equation}
constitute an $SU(2)$ algebra on $H(M)$ \cite{GH}, which inherits spatial rotation
of the underlying four-dimensional theory.

As we already noted in the first section, this fact is
consistent with expectations that intrinsic Higgs states
have something to do with single-center black holes.
When we consider more general quivers, this statement must
be generalized a bit, because additional tree-like component of
the quiver could add extra charge centers, orbiting around the
single-center black hole of the loop, and carry
additional angular momentum. In the Higgs phase description,
this will manifest as a factorization in the ambient
space $X$, say $X=X'\times Y$,  such that
\begin{equation}
H(M)=H(M')\otimes H(Y)\,,
\end{equation}
and our discussion above will apply to $M'\hookrightarrow X'$.

One immediate classification among these arises depending on
whether the dimension $d_k$ is even or odd. When $d_k$ is even,
the cohomology is entirely of even-dimensional, again thanks to
the hyperplane theorem, so that
\begin{equation}
H(M_k)=i^*_{M_k}(H(X_k))\oplus [H^{d_k}(M_k)/i^*_{M_k}(H^{d_k}(X_k))]\,,
\end{equation}
and the Euler number counts the number of ground
states faithfully. When $d_k$ is odd, we have instead
\begin{equation}
H(M_k)=i^*_{M_k}(H(X_k))\oplus H^{d_k}(M_k)\,,
\end{equation}
since $H^{d_k}(X_k)$ is empty. In this case the Euler number
counts the difference between the dimension of pull-back $i^*_{M_k}(H(X_k))$
and that of the middle cohomology $H^{d_k}(M_k)$.

With this in mind, let us consider properties of $i^*_{M_k}(H(X_k))$.
A simple method to recover this is to truncate the well-known
Poincar\'{e} polynomial of $X_k$,
\begin{equation}\label{poincare-poly}
P[{X_k}]=\prod_{i\neq k} (1+x^2+x^4+\cdots+x^{2(a_i-2)})=\sum b_{2l}(X_{k})\cdot x^{2l}
\end{equation}
up to the order $d_k$, and to complete it by
inverting the lower half to the upper half. This
defines the ``Poincar\'{e} polynomial" of the
pulled-back cohomology,
\begin{equation}\label{PXM}
P[i^*_{M_k}(H(X_k))]\equiv
b_{d_k}(X_{k})\cdot x^{d_k}+\sum_{0\le 2l<d_k}  b_{2l}(X_{k})\cdot(x^{2l}+x^{2d_k-2l})\,,
\end{equation}
where, as we noted above, the first term on the right exists only when $d_k$ is even.
It is not difficult to see that $b_{2l}(X_k)$ has three different
regimes, increasing for small $l$, a plateau in the middle, and decreasing
for large $l$.
Thus, counting of $i^*_{M_k}(H(X_k))$ will give different type of
behaviors depending on the value $d_k$ relative to the linking numbers.

A pair of useful numbers, $2N_k$ and $2L_k$, defined, for example for $k=n+1$,
\begin{eqnarray}
&&L_{n+1}= a_1+\cdots +a_n-n-N_{n+1}\ , \\
&&N_{n+1}={\rm max}\left\{a_1-1,a_2-1,\cdots,a_n-1,\left[\frac{a_1+\cdots +a_n-n+1}{2}\right]\right\} \,, \label{N}
\end{eqnarray}
are such that we have
\begin{equation} \label{BettiStructure}
b_0< b_2 <\cdots <b_{2L_{n+1}}=b_{2L_{n+1}+2}=\cdots=b_{2N_{n+1}}> b_{2N_{n+1}+2} >\cdots > b_{2(a_1+\cdots+a_n -n)}\,.
\end{equation}
One obvious category is $d_k \le 2L_k +1$, which in turn
translates to the condition\footnote{This $d_k  \le 2L_k+1$ condition is similar to those for a scaling solution to exist,
$$
a_m < \sum_{i\neq m}^{n+1}a_j\ , \quad \hbox{for } m=1,\dots,n+1\,,
$$
but not quite.}
\begin{equation}
d_{k}\le 2L_{k}+1\quad \leftrightarrow\quad  a_j+(n-3)\le \sum_{i\neq j}^{n+1}a_j \ , \quad j\neq k \ .
\end{equation}
The second category arises when $2L_k +1 <d_k  \le 2N_k +1$, whereby
one finds size of $ i^*_{M_k}H^{2l}(X_k)$ plateaus through a symmetric range
of middle dimensions $2l\sim d_k$. Physically the corresponding ground
states are such that the angular momentum multiplets have some minimum
spin. We believe that, in this case, no intrinsic Higgs states exist at all.

The third apparent  possibility $ 2N_k +1 < d_k $ is never realized,
because it is inconsistent with the Lefschetz
$SU(2)$ symmetry. This can also be seen directly from the numbers as
\begin{equation}
2N_{n+1}\ge  2\left[\frac{a_1+\cdots +a_n-n+1}{2}\right] > d_{n+1}= a_1+a_2+\cdots+a_n-a_{n+1}-n
\end{equation}
with positive $a$'s.
Although this is, from purely geometric viewpoint, a trivial consequence
of the K\"ahlerian property of  $M_k$'s, it does provide a simple consistency
check on the first conjecture that $i^*_{M_k}(H(X_k))$ is a faithful image of the Coulomb
ground states in the Higgs phase. The Coulomb phase states are expected to be
organized  as a sum of angular momentum multiplets of all integer or all
half-integer spins. Degeneracy of a given helicity is then always maximized
at  0 or $\pm1/2$, with non-increasing behavior as we increase the
absolute value of the total helicity.

\section{Analytical Check: 3-Gons}

The simplest example  is a quiver with three nodes ($n=2$). Before we consider
actual quivers, let us first observe that, for $X=\mathbb{CP}^{a-1} \times \mathbb{CP}^{b-1}$
and a complete intersection embedding $i_M: M \hookrightarrow X$ by $c$ F-term constraints, the dimension
\begin{equation}
D \equiv{\rm dim}\,i^*_{M}(H({X})) \ ,
\end{equation}
of the pulled-back cohomology is given by
\begin{equation} \label{obeyed}
D={\cal D}^{(1)}(a,b;c)  \equiv \frac14\left((a+b-c)^2-r\right) \ ,
\end{equation}
with $r=0,1$ for $a+b+c$ even and odd, respectively,
provided that $a, b,$ and $c$ obey
\bea \label{easy}
\nonumber a &\leq& b+c+1 \ , \\
\label{good} b &\leq& c+a+1 \ ,\\
\nonumber c &\leq& a+b+1 \ .
\eea
Otherwise, we have
\bea \label{violated}
D={\cal D}^{(2)}(a,b;c) &\equiv& \begin{cases}
(a-c)\cdot b \   \quad\; {\text{~~if $a>b+c+1$}} \ ,\\
(b-c)\cdot a \  \quad {\text{~~ if $b>c+a+1$}} \ ,\\
0 \quad \quad \quad \quad \quad   {\text{~~~if $c>a+b+1$}} \ ,
\end{cases}
\eea
Note that only one of the three inequalities~(\ref{easy}) can be violated at a time.\footnote{
See Appendix~\ref{AppendixB} for details.
} Here, the superscript for ${\cal D}$  is shown as a
reminders of the class of the corresponding quiver, to which we turn next.

Let us classify the 3-gon quivers with linking numbers $a_1,a_2,a_3$
as follows. We say that the quiver belongs to the first class if the three inequalities,
\begin{eqnarray}\label{firstclass}
a_1 &\le & a_2+a_3+1\ , \nonumber \\
a_2 &\le & a_3+a_1+1\ , \\
a_3 &\le & a_1+a_2+1\ , \nonumber
\end{eqnarray}
are obeyed.
This is reminiscent of the criteria~(\ref{good}) (and hence, the superscript of ${\cal D}^{(1)}$ for cohomology counting in Eq.~({\ref{obeyed})}, for instance).
The second class is defined to be those quivers that violate one of the three inequalities~(\ref{firstclass}).
The violation can happen only for the largest of three linking numbers,
and without loss of generality, we may label it $a_2$. So for the second class
we effectively assume
\begin{equation}
a_2 >  a_3+a_1+1 \ .
\end{equation}
This implies failure of the triangle condition,
which is necessary for a scaling solution to exist. Expectation is that
in such cases the Euler number of Higgs phase equals the Coulomb index,
which we will be verifying along the way.

For simplicity, let us denote, in each branch $k$, the complex dimension of $M_k$ as
\begin{equation}
d_k \equiv {\rm dim}_\IC M_k = a_1 + a_2 + a_3 - 2a_k - 2 \ ,
\end{equation}
and the dimension of the pulled-back cohomology as
\begin{equation}
D_k \equiv {\rm dim}\,i^*_{M_k}(H({X_k})) \ .
\end{equation}
We start with the second class of quivers, and the branch $M_3$ thereof.
Here, $a_3$ plays the role of $c$ in Eq.~(\ref{violated}), so we find \begin{equation}
D_3 =  {\cal D}^{(2)} (a_1, a_2;a_3)=(a_2-a_3)\cdot a_1 \ ,
\end{equation}
which agrees with the known Coulomb phase index in this parameter regime \cite{Denef:2007vg}
\begin{equation}\label{NC}
\Omega_{\rm Coulomb}(a_1, a_2;a_3) =(a_2-a_3)\cdot a_1 \ ,
\end{equation}
as our first conjecture asserts. On the other hand, Denef and Moore
also observed that, under a weaker condition $a_2+1\ge a_3+a_1$,
the Euler number $\chi(M_3)$ of the Higgs phase agrees with
this Coulomb phase index,
\begin{equation}
\Omega_{\rm Coulomb}(a_1, a_2;a_3) =\chi(M_3) \, .
\end{equation}
Taking these two facts together, we find
\begin{equation}
\chi(M_3)-D_3=0\,.
\end{equation}
Computations for $M_1$ proceeds verbatim, with $a_1$ and $a_3$
exchanged, so we have
\begin{equation}
D_1=\Omega_{\rm Coulomb}(a_3, a_2;a_1) =\chi(M_1) \, ,
\end{equation}
and in particular,
\begin{equation}
\chi(M_1)-D_1=0\,.
\end{equation}
The remaining branch $M_2$ is empty, because the number of F-term constraints,
$a_2$, is larger than the complex dimension $d_2=a_1+a_3-2$ of the ambient space.
Therefore, we have
\begin{equation}
\chi(M_2)-D_2=0-0=0\, ,
\end{equation}
which is still consistent with the second conjecture. The final check (for the
first conjecture) is to show that
$\Omega_{\rm Coulomb}(a_1, a_3; a_2)$ also vanishes in this branch; it follows
trivially from (\ref{Coulomb}) with  $a_2>a_1+a_3+1$ taken into account.\footnote{The
corresponding Coulomb phase  should be given by, with $r_{ij}\equiv|\vec x_i-\vec x_j|$
and $\vec x_k\in \mathbb{R}^3$,
$$
{a_1}/{r_{12}}-{a_2}/{r_{23}} >0 \ , \quad
{a_2}/{r_{23}}-{a_3}/{r_{31}} <0
$$
which can be shown to have no solution when $a_2>a_1+a_3$. Thus, the Coulomb phase
is also empty when its counterpart $M_2$ is empty.}

Next, we turn to the quivers in the first class.
Recall that in each of the three branches $M_k$, the pulled-back cohomology $D_k$ is counted by ${\cal D}^{(1)}$ in Eq.~(\ref{obeyed}).
For example, we have
\begin{equation}
\label{moreinter}
D_3= {\cal D}^{(1)}(a_1, a_2;a_3)=\frac14\left((a_1+a_2-a_3)^2-r\right)\,.
\end{equation}
where, again, $r=1,0$ for odd and even $d_3$, respectively.
On the other hand, under a slightly stronger condition of triangle
inequalities, $a_i<a_j+a_k$, the Coulomb index has
been computed as \cite{deBoer:2008zn,Manschot:2011xc}
\begin{equation}\label{scalingC}
\Omega_{\rm Coulomb}(a_1,a_2;a_3)=\frac14\left((a_1+a_2-a_3)^2-r\right)\, ,
\end{equation}
which agrees with Eq.~(\ref{moreinter}).
For $a_i=a_j+a_k$ and $a_i=a_j+a_k+1$, where the Coulomb phase
should be counted by  Eq.~(\ref{NC}) instead of Eq.~(\ref{scalingC}), it so happens that
the two expressions coincide.
Therefore, Eq.~(\ref{scalingC}) works actually  for the entire regime,  $a_i\le a_j+a_k+1$, so that
\begin{equation}
D_3=\Omega_{\rm Coulomb}(a_1,a_2;a_3)\, ,
\end{equation}
again, and generally,
\begin{equation}
D_k=\Omega_{\rm Coulomb}(a_i,a_j;a_k)\ , \quad i \ , j \ ,  k \text{~~distinct}  ,
\end{equation}
which shows that the first conjecture holds true for the quivers in the second class too.
Now, the second conjecture demands that
\begin{equation}\label{identity}
\chi(M_k)-D_k \,,
\end{equation}
is invariant under the choice of $k$.
Starting with Eq.~(\ref{Euler2}) for $\chi(M)$ in three-node
cases~\cite{Denef:2007vg},
\begin{equation}\label{E3}
\chi(M_k)=\frac{a_1a_2 a_3}{a_k}-\int_0^\infty ds\; e^{-s}L_{a_1-1}^1(s)L_{a_2-1}^1(s)L_{a_3-1}^1(s) \ ,
\end{equation}
we find
\begin{eqnarray}
\chi(M_k)- D_k &=& \chi(M_k) - {\cal D}^{(1)}(a_i,a_j;a_k)\nonumber \\
&=&-\frac14\left(a_1^2+a_2^2+a_3^2-2a_1a_2-2a_2a_3-2a_3 a_1-r\right) \nonumber \\
&&-\int ds\;e^{-s}L_{a_1-1}^1(s)L_{a_2-1}^1(s)L_{a_3-1}^1(s) \ ,
\end{eqnarray}
with $r=1,0$ respectively, for odd and even $a_1+a_2+a_3$, which
is clearly invariant under cyclic rotations among $a_1,a_2,a_3$.
That is,  $\chi(M_k)-D_k$ is independent of
 $k=1,2,3$, as conjectured.

So far, we checked the two conjectures against the relevant degeneracies.
However, our analytical check actually goes beyond this. For the type
of quivers in this section, the relevant angular momentum
information of Coulomb phase states can be extracted from the
proposal of Ref.~\cite{Manschot:2011xc}. For the first class
of quivers, one finds exactly one angular momentum multiplet
for each spin $d/2$, $d/2-1$, etc down to $0$ or $1/2$, respectively,
depending on whether $d$ is even or odd. For the second
class, this descending series of angular momentum multiplets
also starts with $d/2$, $d/2-1$, etc,
but is cut off at some positive spin with no angular momentum
multiplet below it. If our first conjecture holds, the same information should
be encoded in the Poincar\'{e} polynomials (\ref{PXM}) of the
pulled-back cohomology in the Higgs phase.
For all 3-gon cases, this comparison has been made and found
to agree with each other completely. In terms of this Poincar\'{e}
polynomial, the two classes distinguish themselves by either
having a monotonically increasing $b_{2l}=l+1$ between $2l=0$
and $2l=2[d/2]$ or reaching a plateau at $2l=2L$ till $2l=2[d/2]$.

This concludes the explicit and analytical demonstration that
the two conjectures hold separately for all possible three-node
cyclic quivers.

\section{Numerical Evidences: 4-Gons, 5-Gons, and 6-Gons }

\begin{table}[h!!!]
{\begin{center}{
\renewcommand{\arraystretch}{1.2}
\begin{tabular}{|c|}\hline
$(a_1, a_2, a_3, a_4)$ \\ \hline \hline
${\begin{array}{c}\\[3.5mm] (2,3,4,11) \\[9.15mm]\end{array}}$ \\ \hline \hline
${\begin{array}{c}\\[3.5mm](2,3,4,9)\\[9.15mm]\end{array}}$ \\ \hline \hline
${\begin{array}{c}\\[3.5mm](4,5,6,7)\\[9.15mm]\end{array}}$ \\ \hline \hline
${\begin{array}{c}\\[3.5mm](5,7,11,13)\\[9.15mm]\end{array}}$ \\ \hline \hline
${\begin{array}{c}\\[3.5mm](11,12,13,14)\\[9.15mm]\end{array}}$ \\ \hline
\end{tabular}
\begin{tabular}{|c||c|c|c|}\hline
$k$&$\chi(M_k)$&$D_k$&$\chi(M_k) - D_k$ \\ \hline \hline
1  & \scriptsize108 & \scriptsize108 & \scriptsize0 \\ \hline
2  & \scriptsize64 & \scriptsize64 & \scriptsize0 \\ \hline
3  & \scriptsize42 & \scriptsize42 & \scriptsize0 \\ \hline
4  & \scriptsize0 & \scriptsize0 & \scriptsize0 \\ \hline \hline

1  & \scriptsize84 & \scriptsize84 & \scriptsize0 \\ \hline
2  & \scriptsize48 & \scriptsize48 & \scriptsize0 \\ \hline
3  & \scriptsize30 & \scriptsize30 & \scriptsize0 \\ \hline
4  & \scriptsize0 & \scriptsize0 &\scriptsize 0 \\ \hline \hline

1  &  \scriptsize-653458&\scriptsize110&\scriptsize-653568  \\ \hline
2  &  \scriptsize-653500&\scriptsize68&\scriptsize-653568  \\ \hline
3  &  \scriptsize-653528&\scriptsize40&\scriptsize-653568  \\ \hline
4  &  \scriptsize-653548&\scriptsize20&\scriptsize-653568  \\ \hline \hline

1  &  \scriptsize-28895778010 & \scriptsize656 & \scriptsize-28895778666  \\ \hline
2  &  \scriptsize-28895778296 &\scriptsize 370 & \scriptsize-28895778666  \\ \hline
3  &  \scriptsize-28895778556 &\scriptsize 110 & \scriptsize-28895778666  \\ \hline
4  & \scriptsize-28895778626 & \scriptsize40 & \scriptsize-28895778666  \\ \hline \hline

1  &  \scriptsize-7025159641580583958&\scriptsize908&\scriptsize-7025159641580584866  \\ \hline
2  &  \scriptsize-7025159641580584140&\scriptsize726&\scriptsize-7025159641580584866  \\ \hline
3  &  \scriptsize-7025159641580584294&\scriptsize572&\scriptsize-7025159641580584866  \\ \hline
4  & \scriptsize-7025159641580584426&\scriptsize440&\scriptsize-7025159641580584866   \\ \hline
\end{tabular}
}\end{center}}
{\caption{\small
Computation of indices $\chi(M_k)$ and $D_k\equiv{\rm dim}\;i^*_{M_k}(H(X_k))$
for five 4-gon quivers, whose edges labeled by $(a_1, a_2, a_3, a_4)$.
For each quiver the indices are computed in four different branches,
but $\chi(M_k)-D_k$  listed in the last column is clearly insensitive to the choice of the branch,
and in particular uniformly zero for the case
that violates geometric $4$-gon condition and thus no scaling solution in the Coulomb phase.
The second example is a marginal case in this sense.}}
\end{table}

\begin{table}[h!!!]
{\begin{center}{
\renewcommand{\arraystretch}{1.2}
\begin{tabular}{|c|}\hline
$(a_1, a_2, a_3, a_4, a_5)$ \\ \hline \hline
${\begin{array}{c}\\[6.6mm] (1,2,3,4,11) \\[12.3mm]\end{array}}$ \\ \hline \hline
${\begin{array}{c}\\[6.6mm](2,3,4,5,6)\\[12.3mm]\end{array}}$ \\ \hline \hline
${\begin{array}{c}\\[6.6mm](2,3,4,7,11)\\[12.3mm]\end{array}}$ \\ \hline \hline
${\begin{array}{c}\\[6.6mm](3,6,9,11,16)\\[12.3mm]\end{array}}$ \\ \hline
\end{tabular}
\begin{tabular}{|c||c|c|c|}\hline
$k$&$\chi(M_k)$&$D_k$&$\chi(M_k) - D_k$ \\ \hline \hline
1  & \scriptsize240&\scriptsize240&\scriptsize0\\ \hline
2  & \scriptsize108&\scriptsize108&\scriptsize0\\ \hline
3  & \scriptsize64&\scriptsize64&\scriptsize0\\ \hline
4  & \scriptsize42&\scriptsize42&\scriptsize0\\ \hline
5  & \scriptsize0&\scriptsize0&\scriptsize0\\ \hline \hline

1  & \scriptsize173100&\scriptsize259&\scriptsize172841\\ \hline
2  & \scriptsize172980&\scriptsize139&\scriptsize172841\\ \hline
3  & \scriptsize172920&\scriptsize79&\scriptsize172841\\ \hline
4  & \scriptsize172884&\scriptsize43&\scriptsize172841\\ \hline
5  & \scriptsize172860&\scriptsize19&\scriptsize172841\\ \hline \hline

1  & \scriptsize3650745&\scriptsize951&\scriptsize3649794\\ \hline
2  & \scriptsize3650360&\scriptsize566&\scriptsize3649794\\ \hline
3  & \scriptsize3650052&\scriptsize258&\scriptsize3649794\\ \hline
4  & \scriptsize3649920&\scriptsize126&\scriptsize3649794\\ \hline
5  & \scriptsize3649800&\scriptsize6&\scriptsize3649794\\ \hline \hline

1 & \scriptsize-10110744325279026 & \scriptsize7852 & \scriptsize-10110744325286878 \\ \hline
2 & \scriptsize-10110744325283778 & \scriptsize3100 & \scriptsize-10110744325286878 \\ \hline
3 & \scriptsize-10110744325285362 & \scriptsize1516 & \scriptsize-10110744325286878 \\ \hline
4 & \scriptsize-10110744325285938 & \scriptsize940 &\scriptsize -10110744325286878 \\ \hline
5 & \scriptsize-10110744325286748 & \scriptsize130 & \scriptsize-10110744325286878 \\ \hline
\end{tabular}

}\end{center}}
{\caption{\small Computation of indices $\chi(M_k)$ and $D_k\equiv{\rm dim}\;i^*_{M_k}(H(X_k))$
for four 5-gon quivers, whose edges labeled by $(a_1, a_2, a_3, a_4, a_5)$.
For each quiver the indices are computed in five different branches,
but $\chi(M_k)-D_k$ listed in the last column is clearly insensitive
to the choice of the branch, and in particular uniformly zero for the case
that violates geometric $5$-gon condition. }}
\end{table}

\begin{table}[h!!!]
{\begin{center}{
\renewcommand{\arraystretch}{1.2}
\begin{tabular}{|c|}\hline
$(a_1, a_2, a_3, a_4, a_5, a_6)$ \\ \hline \hline
${\begin{array}{c}\\[9.7mm] (1,2,4,5,6,19) \\[15.45mm]\end{array}}$ \\ \hline \hline
${\begin{array}{c}\\[9.7mm](3,4,5,6,7,9)\\[15.45mm]\end{array}}$ \\ \hline \hline
${\begin{array}{c}\\[9.7mm](2,3,5,7,11,13)\\[15.45mm]\end{array}}$ \\ \hline
\end{tabular}
\begin{tabular}{|c||c|c|c|}\hline
$k$&$\chi(M_k)$&$D_k$&$\chi(M_k) - D_k$ \\ \hline \hline
1 & \scriptsize4320 & \scriptsize4320 & \scriptsize0\\ \hline
2 & \scriptsize2040 & \scriptsize2040 & \scriptsize0\\ \hline
3 & \scriptsize900 & \scriptsize900 & \scriptsize0\\ \hline
4 & \scriptsize672 & \scriptsize672 & \scriptsize0\\ \hline
5 & \scriptsize520 & \scriptsize520 & \scriptsize0\\ \hline
6 & \scriptsize0 & \scriptsize0 & \scriptsize0\\ \hline \hline

1 & \scriptsize-189808421214888 &\scriptsize 5488 &\scriptsize -189808421220376\\ \hline
2 & \scriptsize-189808421216778 &\scriptsize 3598 &\scriptsize -189808421220376\\ \hline
3 & \scriptsize-189808421217912 &\scriptsize 2464 &\scriptsize -189808421220376\\ \hline
4 & \scriptsize-189808421218668 &\scriptsize 1708 &\scriptsize -189808421220376\\ \hline
5 & \scriptsize-189808421219208 &\scriptsize 1168 &\scriptsize -189808421220376\\ \hline
6 & \scriptsize-189808421219928 &\scriptsize 448 &\scriptsize -189808421220376\\ \hline \hline

1 & \scriptsize1513169068553549 & \scriptsize12979 & \scriptsize1513169068540570\\ \hline
2 & \scriptsize1513169068548544 & \scriptsize7974 & \scriptsize1513169068540570\\ \hline
3 & \scriptsize1513169068544540 & \scriptsize3970 & \scriptsize1513169068540570\\ \hline
4 & \scriptsize1513169068542824 & \scriptsize2254 & \scriptsize1513169068540570\\ \hline
5 & \scriptsize1513169068541264 & \scriptsize694 & \scriptsize1513169068540570\\ \hline
6 & \scriptsize1513169068540844 & \scriptsize274 & \scriptsize1513169068540570\\ \hline

\end{tabular}

}\end{center}}
{\caption{\small Computation of indices $\chi(M_k)$ and $D_k\equiv{\rm dim}\;i^*_{M_k}(H(X_k))$
for three 6-gon quivers, whose edges labeled by $(a_1, a_2, a_3, a_4, a_5, a_6)$.
For each quiver the indices are computed in six different branches, but $\chi(M_k)-D_k$
listed in the last column is clearly insensitive to the choice of the branch, and
in particular uniformly zero for the case that violates geometric $6$-gon condition.}}
\end{table}

Let us now turn to more involved examples with many nodes.
As before, we will consider distinct branches, labeled by
$k=1,2,\dots,n+1$, and compute the indices $\chi(M_k)$
and $D_k\equiv\chi(i^*_{M_k}(H(X_k)))=
{\rm dim}\;i^*_{M_k}(H(X_k))$ individually. We show then that the
intrinsic Higgs states, counted by the difference, $\chi(M_k)-D_k$,
are invariant as we move from one branch to another. Tables 1, 2 and 3
each illustrate this, for 4-gon, 5-gon and 6-gon quivers, respectively.

While we do not have general formulae for $\Omega_{\rm Coulomb}$,
there are a couple of important checks we can perform.
First, general expectation from physical considerations
is that, when no scaling solution exists in the Coulomb phase, Higgs and
Coulomb should agree with each other \cite{Denef:2002ru,Sen:2011aa}.
When the geometric $(n+1)$-gon condition fails, the first conjecture
combined with the expectation $\chi=\Omega_{\rm Coulomb}$ implies that
$$\chi(M)-{\rm dim}\;i^*_M(H(X)) =0 \ ,$$
regardless of the branch. This can be seen to hold with the example
at the top of each table.

More generally, 
we show by explicit examples that this difference which can be
typically very large is independent of the choice of branch,
consistent with the second conjecture.
As noted in the first section, the two conjectures  are
physically linked to each other as long as the wall-crossing
behavior is entirely captured by the Coulomb phase ground states.
In this indirect sense, we are effectively testing validity
of the first conjecture as well.

\section{Comments}

In this note we conjectured that the difference between Coulomb
phase and Higgs phase indices can be given a purely Higgs
phase characterization, namely the difference between the
full cohomology of Higgs phase itself $M$ and the pull-back to $M$ of
cohomology of the ambient D-term vacuum manifold $X$.
The statement that Coulomb phase states
correspond to the pulled-back cohomology of
ambient D-term toric variety is well motivated by the fact
that the two sides agree when no loops are present and thus
no F-term constraints arise either. Introducing loops and thus
F-terms tends to make Higgs phase relatively more complicated, with the
known result that the intrinsic Higgs states begin to show and in fact
can be exponentially more numerous. In some crude sense, the Coulomb
states are already known to D-term-induced ambient space
in the Higgs side, to which F-terms add more states.

Since
wall-crossing physics arises inherently from Coulombic physics
of multi-centered nature, this leads to the second conjecture
that the difference is in fact an invariant of the quiver itself.
Traditional
invariants, such as $\Omega_{\rm Coulomb}$ that directly enters
the wall-crossing formulae, and the Euler number $\chi$ of
Higgs phase, experience wall-crossing and thus change discontinuously
as we deform FI constants. The difference
\begin{equation}
\chi-\Omega_{\rm Coulomb}
\end{equation}
according to our first conjecture, becomes a geometric object
entirely of Higgs phase,
\begin{equation}
\chi(M_k)-\chi(i^*_{M_k}H(X_k))\, ,
\end{equation}
which, under the second conjecture, is an invariant of the quiver itself.
This is one very intriguing  consequence of this work.
Equality of this number among different Higgs branches
implies an invariant of the quiver diagram itself,
rather than individual branches. This aspect may be
further explored  in purely mathematical terms also.

We proved the conjectures for the simplest case of all $3$-gon quivers, 
for which ground states in Coulomb and Higgs phases
are catalogued explicitly. We also checked the conjectures
numerically for a large number of $4$-gon, $5$-gon and $6$-gon quivers,
and presented a few typical examples in the note. More thorough
and analytical treatments  of the conjectures, including further
proofs and consistency checks, will appear elsewhere \cite{LWY}.

\vskip 5mm

As this work was drawing to conclusion, Ref.~\cite{Bena:2012hf}
appeared with some partially overlapping results.

\vskip 1cm
\centerline{\bf\large Acknowledgments}
\vskip 5mm\noindent
We are grateful to Ashoke Sen for useful discussions.
This work is supported by the National Research Foundation of Korea
(NRF) funded by the Ministry of Education, Science and Technology
with grant number 2010-0013526.
\appendix


\section{Lefschetz Hyperplane Theorem} \label{AppendixA}

Let $X$ be a complex compact manifold and let $\lambda$ be a
holomorphic section of a positive holomorphic line bundle $\CL$ over $X$.
The section $\lambda$ can be thought of as a defining polynomial
for the hypersurface $M=\lambda^{-1}(0) \subset X$.
We then have, from Lefschetz hyperplane theorem~\cite{GH}, that
the natural map $H^p(X,\IZ) \rightarrow H^p(M, \IZ)$ is an
isomorphism for $ p  < {\rm dim}_\IC M$ and is an injection for $p={\rm dim}_\IC M$.

A straight-forward generalisation to complete intersection cases
can be made by iteration:
Let $\lambda_1, \lambda_2, \dots, \lambda_m$ be the sections
of positive line bundles $\CL_1$, $\CL_2$, $\dots$, $\CL_m$ over $X$,
and let $M^{(r)}$, for $r \leq m$, be the complete intersections of the
first $r$ hypersurfaces $\lambda_1^{-1}(0), \dots, \lambda_r^{-1}(0)$ to $X$.
We then have the following chain of hypersurfaces
\begin{equation}
M^{(m)} \subset M^{(m-1)} \subset \cdots \subset M^{(1)} \subset M^{(0)} \equiv X \ ,
\end{equation}
where the ambient space $X$ is denoted by $M^{(0)}$.
We begin with the first line bundle $\CL_1$ and consider
the hypersurface $M^{(1)} \subset X$.
Since the $r$-th line bundle $\CL_r$ is positive over $X$,
it is also positive over the subspace $M^{({r-1})}$, which
is the complete intersection of the previous $r-1$ hypersurfaces.
Now, by applying Lefschetz hyperplane theorem to the
hypersurface $M^{(r)} \subset M^{(r-1)}$, we have the isomorphisms
\begin{equation}
H^p(M^{(r-1)}, \IZ) \simeq H^p(M^{(r)} , \IZ)\ , ~ p<{\rm dim}_\IC M^{(r)} \ ,
\end{equation}
which leads, when iteratively applied, to
\begin{equation}
H^p(X , \IZ) \simeq H^p(M^{(m)}, \IZ) \ , ~p<{\rm dim}_\IC M^{(m)} \ ,
\end{equation}
for all the lower-half cohomologies but for the middle one.
Similarly, we also see that the natural map $H^p(X, \IZ) \to H^p(M^{(m)}, \IZ)$
for $p={\rm dim}_\IC M^{(m)}$, to the middle cohomology of $M^{(m)}$, is injective.

\section{An Exercise in Cohomology Counting} \label{AppendixB}
Let us recall that the Poincar\'{e} polynomial for $X=\mathbb{CP}^{a-1}\times \mathbb{CP}^{b-1}$, with $b\ge a$, is given by
\begin{equation}\label{3-poincare}
P[X] =\underbrace{1+ \cdots + a \cdot x^{2(a-1)}}_{1, ~2, ~\cdots~, ~a}
+  \underbrace{a \cdot x^{2a} + \cdots + a \cdot x^{2(b-1)}}_{a, ~a,~\cdots~,~a}+
\underbrace{(a-1) \cdot x^{2b} + \cdots + x^{2(a+b-2)}}_{a-1, ~a-2, ~\cdots~, ~1}  \ ,
\end{equation}
which naturally splits into three parts, in accord with the general structure~(\ref{BettiStructure}), with $2L+1\equiv2a-1$.
We embed a complete intersection $M$ using $c$ F-term constraints,
and ask what the pulled-back cohomology looks like. The answer should
depend on where the $x^d$-term, with $d \equiv {\rm dim}_\IC M = a+b-c-2$,  sits
in $P[X]$ relative to the power $2L+1$.

When $d>2L+1$ (or equivalently, $b > a+c+1$), the $x^d$-term sits in the middle plateau of $P[X]$, and
\bea \nonumber
 {\rm dim}\,i^*_{M}(H({X})) -{\rm dim}\,i^*_{M}(H^{d}(X))
 &=& 2 \cdot \left( (1+ 2 + \cdots + a) + a \cdot \left[\frac{d  -2a+1}{2}\right] \right)\\
 & = &
\nonumber  \begin{cases}(b-c) \cdot a \quad&\mbox{if } d  {~\text{odd}}\,,\\
(b-c-1) \cdot a\quad&\mbox{if } d  {~\text{even}}\,.
\end{cases}
\eea
Since ${\rm dim}\,i^*_{M }(H^{d }(X ))= 0$, $a$,
for $d $ odd and even, respectively, we have
\begin{equation}
{\rm dim}\,i^*_{M }(H({X })) = (b-c)\cdot a  \ ,
\end{equation}
independent of the parity of $d $.
On the other hand, when $d\le 2L+1$ (or equivalently, $b \le a+c+1$), the $x^d$-term now sits in the first part of $P[X]$, and the cohomology counting in this case is given by
\bea \nonumber
 {\rm dim}\,i^*_{M }(H({X })) -{\rm dim}\,i^*_{M }(H^{d }(X ))&=& 2
 \cdot \left( 1+2+\cdots+\left[ \frac{d +1}{2}\right]\right) \\
 & = &
\nonumber \begin{cases}\frac14\left((a+b-c)^2-1\right)\quad&\mbox{if } d  {~\text{odd}}\,,\\
\frac14(a+b-c)^2 - (\frac d2 +1) \quad&\mbox{if } d  {~\text{even}}\,.
\end{cases}
\eea
Since ${\rm dim}\,i^*_{M }(H^{d }(X ))=0, d/2 +1$, for $d $ odd and even, respectively, we have
\begin{equation}\label{tD}
{\rm dim}\,i^*_{M }(H({X })) = \frac14\left((a+b-c)^2-r\right) \ ,
\end{equation}
where $r=1, 0$ for $d $ odd and even, respectively.

\end{document}